\documentclass[epj]{svjour}
\usepackage{graphicx}
\usepackage{times}
\usepackage{amssymb}

\begin{document}
\title{Prospects for precision measurements of atomic helium using direct frequency comb spectroscopy}
\author{E. E. Eyler\inst{1}
\and D. E. Chieda\inst{1}
\and Matthew C. Stowe\inst{2}
\and Michael J. Thorpe\inst{2}
\and T. R. Schibli\inst{2}
\and Jun Ye\inst{2}
}                     
%
%
\institute{Physics Department, University of Connecticut, Storrs, CT 06269, USA, \email{eyler@phys.uconn.edu}
 \and JILA, National Institute of Standards and Technology and University of Colorado Department of Physics, University of Colorado, Boulder, CO 80309-0440, USA}

\date{Received: date / Revised version: date}
%
\abstract{ We analyze several possibilities for precisely measuring electronic transitions in
atomic helium by the direct use of phase-stabilized femtosecond frequency combs.  Because the comb
is self-calibrating and can be shifted into the ultraviolet spectral region via harmonic
generation, it offers the prospect of greatly improved accuracy for UV and far-UV transitions.  To
take advantage of this accuracy an ultracold helium sample is needed.  For measurements of the
triplet spectrum a magneto-optical trap (MOT) can be used to cool and trap metastable $2\,^3S$
state atoms. We analyze schemes for measuring the two-photon $2\,^3S \rightarrow 4\,^3S$ interval,
and for resonant two-photon excitation to high Rydberg states, $2\,^3S \rightarrow 3\,^3P
\rightarrow n^3S,D$. We also analyze experiments on the singlet-state spectrum. To accomplish this
we propose schemes for producing and trapping ultracold helium in the $1\,^1S$ or $2\,^1S$ state
via intercombination transitions. A particularly intriguing scenario is the possibility of
measuring the $1\,^1S \rightarrow 2\,^1S$ transition with extremely high accuracy by use of
two-photon excitation in a magic wavelength trap that operates identically for both states.  We
predict a ``triple magic wavelength'' at 412 nm that could facilitate numerous experiments on
trapped helium atoms, because here the polarizabilities of the $1\,^1S$, $2\,^1S$ and $2\,^3S$
states are all similar, small, and positive.
} 
\maketitle
\section{Introduction}
\label{sec:intro}

As the simplest two-electron atom, helium has long been a favorite testing ground for fundamental
two-electron QED theory and new techniques in atomic physics, both experimental and theoretical. A
good summary of the overall status of its energy level spectrum is provided in a recent
publication by Morton, Wu, and Drake \cite{Drake05}.  Overall, experiment and theory are in
reasonable agreement, although for many energy levels the theoretical uncertainties are smaller
than the experimental ones.  In a few important cases significant discrepancies arise, most
notably for the fine structure of the $2\,^3P$ state
\cite{Shiner00,Hessels00,Hessels01,Inguscio05,Gabrielse05,Drake02,Pachucki03,Pachucki06}.  Of
special interest is the ionization energy (IE) of the ground $1\,^1S_0$ state of $^4$He, which
provides a sensitive test of QED corrections for a two-electron system.  It is known
experimentally to about 45 MHz, a fractional accuracy of $9 \times 10^{-9}$, and the most recent
theoretical calculation has surpassed this with an estimated uncertainty of 36 MHz
\cite{Pachucki06b}.  The ground state poses unique experimental challenges because of the huge gap
of about 20 eV to the lowest excited states, and the most recent measurements
\cite{Bergeson98,Bergeson00,Eikema97} are starting to approach the fundamental limits of
nanosecond far-UV laser technology.  The best accuracy so far attained in any system with
nanosecond lasers is about one part in $10^9$ for two-photon transitions near 200 nm
\cite{Ubachs06}, and about six parts in $10^9$ for one-photon transitions below 100 nm
\cite{Trickl07}. Even for transitions between excited states of helium in the visible and near-UV
spectral region, the accuracy of recent experimental measurements cannot easily be improved using
conventional laser spectroscopy with room-temperature samples.

In this paper we propose several measurement approaches and evaluate their feasibility.  All would
use direct frequency comb spectroscopy (DFCS), together with an ultracold $^4$He sample, to
greatly improve the accuracy of selected UV and far-UV transitions. Because narrow natural
linewidths are required for high-resolution spectroscopy, all of the experiments involve either
the ground $1\,^1S_0$ state or one of the two metastable states, $2\,^1S_0$ and $2\,^3S_1$, which
have lifetimes of 1/51 s and 8000 s, respectively \cite{Lin77,Drake71}.

The DFCS technique, which takes full advantage of the extremely accurate pulse-to-pulse coherence
of a phase-stabilized femtosecond frequency comb \cite{Ye_review}, combines the power of high
spectral resolution and broad spectral coverage in spectroscopy. The self-calibrating capability
of the comb permits accurate determination of atomic structure in alkali and alkaline earth atoms
\cite{Marian04,Marian05,Gerginov05,Pichler05,Hansch07,Hollberg07}, as well as the investigation of
multiple molecular transitions \cite{Ye06}.  Furthermore, high-resolution quantum control can be
achieved via spectral manipulation of frequency combs \cite{Ye06b,Ye07}.

Here we discuss in detail some of the considerations that would arise in applying DFCS with continuous pulse trains to precision UV spectroscopy.  For the case of near-resonant two-photon excitation we perform numerical modeling that predicts unusual lineshapes caused by power-dependent shifts, saturation, and two-photon excitation in a counterpropagating pulse geometry with dissimilar wavelengths.  In Section \ref{sec:triplet} of this paper we discuss possible applications of this experimental scheme and others to measuring the triplet spectrum. These experiments would serve a dual purpose: in addition to the intrinsic physical interest of improving the helium energy levels, the excited-states spectrum of helium can serve as an excellent testbed for developing and evaluating
DFCS spectroscopy.

In Section \ref{sec:Intercombination} the emphasis shifts towards the singlet spectrum.  We propose methods for efficiently producing ultracold helium atoms either in the singlet ground state or in the $2\,^1S_0$ state, and for optically trapping them.  We explore prospects for DFCS measurements of singlet-triplet intercombination transitions between excited states.  Finally, in Section \ref{sec:Singlets} we explore prospects for using DFCS to achieve self-calibrating measurements of the ground-state energy.  This would be accomplished by exploiting the recently developed ability to generate far-UV harmonics of the comb that preserve most of the pulse-to-pulse phase coherence \cite{Ye05,Hansch05,Eikema05,Eikema06}, a technique that could bring to the far-UV region the same spectacular improvements in frequency metrology that the comb has enabled in the visible region. A particularly exciting possibility is to use the comb harmonics centered at 120 nm in conjunction with ``magic-wavelength" optical trapping to excite the $1\,^1S \rightarrow 2\,^1S$ transition with a linewidth as small as 300 Hz.

\section{DFCS and UV frequency combs}
\label{sec:DFCS}

\begin{figure}
\centering \vskip 0 mm
\includegraphics[width=0.95\linewidth]{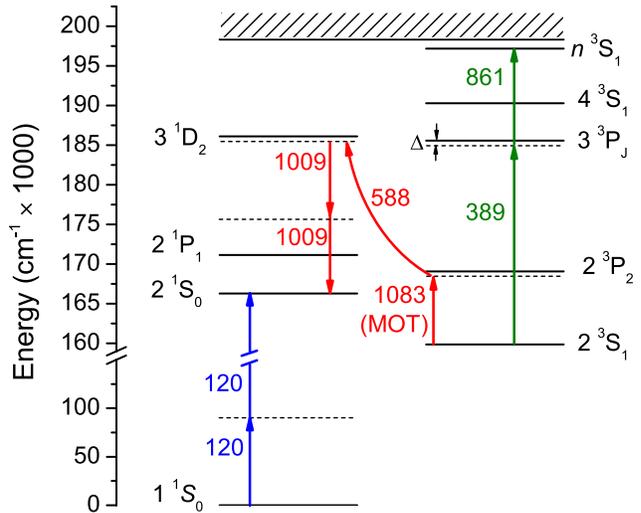}
\caption{\protect\label{He_Levels} (color online) Selected energy levels of helium, showing three of the processes described here, with wavelengths in nm: (1) DFCS of the far-UV $1\,^1S \rightarrow 2\,^1S$ transition (left, blue). (2) Production of ultracold $2\,^1S$ state atoms by a four-step Raman transfer using an intercombination transition to the $3\,^1D_2$ state (center, red). (3) Near-resonant DFCS of the $2\,^3S \rightarrow 10\,^3D$ transition (right, green).}
\end{figure}

The concept of using a coherent pulse train for high-resolution spectroscopy dates back to the 1970s \cite{Hansch77,Hansch78}, but it was only in 2004 that high-quality DFCS results were obtained using the modern generation of phase-stabilized femtosecond frequency combs \cite{Marian04}.  These systems exhibit remarkable pulse-to-pulse phase coherence and frequency stabilities better than one part in $10^{14}$. Both one- and two-photon transitions have been demonstrated in DFCS between low-lying states of alkali atoms \cite{Marian04,Marian05,Gerginov05,Pichler05}.  Although a typical frequency comb has $10^4-10^5$ frequency-domain teeth spanning a wavelength range from 750-850 nm or more, usually only a single tooth of the comb is useful for exciting one-photon transitions (an exception is the case of room-temperature Doppler-broadened spectra, for which several teeth can contribute to coherent velocity-selection effects \cite{Pichler05,Pichler07}). In the two-photon case there are two quite distinct limiting cases. If there is a near-resonant intermediate state as sketched in the right-hand portion of Fig. \ref{He_Levels}, then the two comb frequencies closest to the intermediate-state resonance provide the dominant contribution to the transition amplitude as the comb spacing is normally much larger than a typical atomic transition linewidth. Furthermore, contributions from other nearby comb components tend to cancel in symmetrically detuned pairs due to the opposite phase between them on either side of the resonance, although this can be altered by applying spectral phase shaping to the frequency comb \cite{Meshulach99}. A near resonance can always be arranged, because the frequency of the $n^{\textrm{th}}$ mode of the comb depends on two parameters, $\nu_n=f_0+nf_r$, where the pulse repetition frequency $f_r$ and the carrier envelope offset frequency $f_o$ can be adjusted independently.  The other limiting case occurs if the entire comb is far from intermediate-state resonance.  Then many different pairs of teeth can add to give the same total transition energy, and in favorable situations all can contribute constructively \cite{Felinto04,Kielpinski06}.

Some of the measurements proposed here require UV or far-UV wavelengths.  It is not difficult to
produce wavelengths down to 200 nm by generating the second or third harmonic in nonlinear
crystals such as $\beta$ barium borate (BBO).  Although the limited phase-matching range
spectrally narrows the resulting UV comb, this is usually an advantage, because it reduces the
likelihood that when a given atomic transition is studied by scanning $f_r$, a distant comb tooth
will inadvertently excite an unrelated resonance.  For wavelengths below 200 nm an exciting new
approach is to employ harmonic generation in an external optical resonator \cite{Ye05,Hansch05}.
Continuous pulse trains involving harmonics up to the fifteenth order have been produced by this
method, although the available power drops rapidly as the harmonic order increases.  An
alternative approach is to amplify a short segment of the comb with a regenerative amplifier prior
to harmonic generation \cite{Eikema05,Eikema06}, producing a sequence of a small number of far-UV
pulses with much higher power.  This approach has already been used to observe Ramsey-type fringes
in the spectrum of xenon at 125 nm \cite{Eikema06}.  Compared with the continuous pulse trains
that we analyze here, these short pulse sequences should make the observation of two-photon helium
transitions considerably easier, although at the cost of increased bandwidth and the need to
correct for phase shifts. In the measurements proposed below we assume if not specified otherwise
that a visible comb is available with about 1~W of average power and a repetition frequency of 500
MHz (chosen towards the high end of the readily attainable range, to minimize the likelihood of
accidentally exciting multiple resonances at the same $f_0$ and $f_r$). We assume that a
second-harmonic conversion efficiency of 25\% can be achieved, and that for the seventh harmonic
near 120 nm an efficiency of $2 \times 10^{-6}$ is achievable by designing an optimized
phase-matched configuration. This considerably exceeds the far-UV power demonstrated so far for
continuous pulse trains, but is within the projected range of attainable efficiencies
\cite{Hansch05}.  A resonant enhancement cavity \cite{Ye06} might also be used to enhance the
far-UV power.

To take advantage of the high resolution of the frequency comb an ultracold helium sample is needed.  Otherwise, even if Doppler-free two-photon spectroscopy were utilized, the thermal velocity would severely limit the transition linewidth due to the transit time of the atoms through a focused laser beam.  In the proposed measurements below we expect that a magneto-optical trap (MOT) operating on the $2\,^3S_1 - 2\,^3P_2$ transition is available with fairly typical properties \cite{Vassen99}: $\gtrsim 10^8$ atoms in the triplet metastable state at a density of $4 \times 10^9$ cm$^{-3}$, with a temperature of 1 mK that can be reduced to  as low as 100 $\mu$K using a molasses cooling step.  A magnetic field ramp could also be added to increase the density \cite{Dos Santos 02}.  In Sections \ref{sec:Intercombination} and \ref{sec:Singlets} we also discuss optical trapping, including a ``magic wavelength" trap for atoms in the $1\,^1S$ and $2\,^1S$ states.

\section{Triplet spectra}
\label{sec:triplet}

While it is the singlet spectrum that could benefit most dramatically from recent developments in
DFCS, there are also several worthwhile measurements in the triplet system.  For many purposes the
triplet metastable state of helium can be regarded as a second ground state, one far more
accessible to experimenters than the $1\,^1S$ state, so it has been the focus of much recent
activity
\cite{Shiner00,Hessels00,Hessels01,Inguscio05,Gabrielse05,Drake02,Pachucki03,Pachucki06,Pachucki06c}.
Continued advances in the theory of two-electron quantum electrodynamic (QED) and relativistic
corrections have been spurred by improved experimental results, and particularly by the prospect
of obtaining an improved value for the fine-structure constant from newly obtained and extremely
accurate measurements of transitions between the triplet $n$=2 and $n$=3 states
\cite{Gabrielse05}, assuming that present inconsistencies in the theoretical work can be resolved
\cite{Drake02,Pachucki03}.  The accuracy of current experimental work has become severely
constrained by the motion of room-temperature atoms in free space, and major improvements will
require trapped-atom experiments.  The ``magic wavelength" optical lattice traps described in
Sections \ref{sec:Intercombination_trapping} and \ref{sec:Singlets_1s2s} would provide non-perturbative trapping for the $2\,^3S \rightarrow
2\,^1S$ and $1\,^1S \rightarrow 2\,^1S$ transitions.  We note that a magic wavelength should also
exist near 698 nm for the $2\,^3S \rightarrow 2\,^3P$ transition, although the polarizability is
negative.

Surprisingly there appear to be no highly accurate determinations of transitions from the $2\,^3S$
state to high-$n$ Rydberg states.  As a result, the widely cited value for the ionization energy
(IE) of the metastable state is not a conventional experimental determination based on
extrapolation of a Rydberg series, but instead a hybrid result obtained by combining an extremely
accurate measurement of the $2\,^3S \rightarrow 3\,^3D_{1,2,3}$ transition \cite{Dorrer97} with
theoretical predictions of QED and relativistic corrections for the $3\,^3D_J$ state.  The
accuracy is cited as 0.06 MHz in Ref. \cite{Drake05}, but this IE as well as the entire triplet
energy spectrum was shifted by almost two standard deviations since a similar survey article was
published in 1998 \cite{Drake98}, because the theory for the $3\,^3D$ state was revised in the
interim. If one does not rely on QED calculations, the IE of the $2\,^3S$ state is probably
accurate only to about 15 MHz \cite{Martin87}. This striking lack of modern experimental data
would be remedied by the experiments proposed here.

\subsection{Two-photon far-off-resonance excitation of triplet Rydberg states}
\label{sec:triplet_two-photon}

Transitions from the $2\,^3S$ state to low-lying triplet excited states present an excellent example of two-photon DFCS in the limit where the first photon is far from resonance with any intermediate state.  A particularly good candidate is the $2\,^3S \rightarrow 4\,^3S$ transition, which can be excited by two equal photons at 657 nm, and has so far been measured only to 2.4 MHz accuracy \cite{Hlousek83}.  In a counterpropagating-beam geometry the excitation is Doppler-free and has a natural linewidth of 2.5 MHz.  We detect the $4\,^3S$ state by photoionizing the atoms using surplus 1083 nm MOT trapping light.   For 1 W focused to a 35 $\mu$m gaussian 1/$e^2$ radius, the ionization probability is roughly 1-2\% during the $4\,^3S$ lifetime of $\tau_{4s} = 58$~ns, and the atoms that are not ionized return to the $2\,^3S$ state and can be recycled.

Because the 657 nm wavelength is easily produced by cw lasers, it is possible to measure this
transition either by DFCS or by conventional cw two-photon spectroscopy, with the frequency comb
used only for calibration.  Thus a direct comparison of the two methods is possible. Given the 2.5
MHz natural linewidth, an accuracy better than 0.05 MHz can be anticipated, greatly exceeding the
present theoretical uncertainty of 0.7 MHz.

We make an order-of-magnitude estimate of the transition probability by considering only the $3\,^2P$ intermediate state in a perturbation theory calculation of the two-photon transition amplitude. Because the limiting linewidth is the natural linewidth, we expect a steady-state excitation rate that will not be affected by transient effects such as the buildup of internal coherence \cite{Felinto04}, and the required DFCS laser power will be closely comparable to that for a cw laser with the same detuning.  This is because many different pairs of frequency-domain comb teeth are simultaneously resonant with the two-photon transition, and for transform limited pulses the transition amplitudes can add constructively as already mentioned in Section \ref{sec:DFCS}.  For a given total laser power distributed into a comb with $N$ teeth, the two-photon probability for a single pair of teeth used alone would be reduced by $N^2$. However, the coherent contribution of $N$ pairs of teeth yields a canceling enhancement factor of $N^2$.  The same point has been emphasized by Kielpinski \cite{Kielpinski06} in the context of laser cooling. In summary, \textit{for two-photon excitation far from intermediate-state resonance, only the time-averaged irradiance of the frequency comb is important}; so long as the two-photon resonance condition is met, the signal is nearly independent
of $f_r$ and the bandwidth.

Making these approximations, the effective Rabi frequency at exact two-photon resonance for either
DFCS or cw excitation is given for a laser of time-averaged irradiance $I$ by \begin{equation}
\label{2-photon_Rabi} \Omega_{2\gamma} = \frac{{D_{4s-2p} D_{2p-2s}}}{{\varepsilon_0 c\hbar ^2
}}\frac{I}{\Delta }. \end{equation} The intermediate-state detuning is $\Delta= 1.13\times
10^{15}$~s$^{-1}$ and the electric dipole matrix elements can be determined from the known $A$
coefficients \cite{NIST}, giving $D_{4s-2p} = 3.6 \times 10^{-30}$ C$\cdot$m and $D_{2p-2s}=-2.1
\times 10^{-29}$ C$\cdot$m.  We assume that the filtered comb has a time-averaged power of 50 mW,
and that the counterpropagating laser beams are focused to a radius of $\omega_0=35\ \mu$m.  For
these conditions, $\Omega_{2\gamma}=1.2 \times 10^5$~s$^{-1}$ and the resulting $2\,^3S
\rightarrow 4\,^3S$ transition rate per atom at exact two-photon resonance is
\begin{equation}
R = \Omega_{2\gamma} ^2 \tau_{4s} = 930 {\rm{~s}}^{-1}.
\end{equation}

This is sufficient to yield an easily detectable signal, and could be further enhanced by tighter
focusing.  At these relatively low average irradiances only small ac Stark shifts are expected.
From the dynamic polarizability of the $2\,^3S$ state \cite{Chen95} a shift of +38 kHz is found,
and a somewhat smaller shift of the $4\,^3S$ state in the opposite direction is expected.  The residual
Doppler width is just 1.5 MHz at $T=100\ \mu K$ for counterpropagating beams, assuming the comb is
slightly spectrally filtered to restrict its bandwidth to 20\%.  Thus the measurement accuracy
will be limited primarily by the 2.5 MHz natural linewidth and the need to correct for trap shifts
in the MOT (a 100 $\mu$K trap depth is equivalent to 2 MHz). These trap shifts could be eliminated by
turning the trap off during the measurement.

\subsection{Numerical simulation of near-resonant DFCS transitions to $n\,^3S$ and $n\,^3D$ Rydberg states}
\label{sec:triplet_Rydberg}

We have used a density matrix based model to investigate the expected lineshapes of two-photon
transitions to high-$n$ triplet Rydberg states, excited by a femtosecond comb via a resonant
intermediate state.  An example is shown in the right-hand portion of Fig. 1.  The transition used
for detailed modeling is $2\,^3S_1 \rightarrow 3\,^3P_0 \rightarrow 40\,^3S_1$, for which the
necessary wavelengths are 789 nm and 389 nm, easily accessible by using the comb and its second
harmonic. The Liouville-von Neumann equation for the density matrix of a closed three-level system
is solved numerically, and the natural linewidths of 1.6 MHz for $3\,^3P_0$ and 3.25 MHz for
$40\,^3S_1$ are included via phenomenological dephasing terms. In this time-domain model, the mode
structure of many thousands of comb components emerges naturally from interference among the
phase-coherent pulses in the femtosecond pulse train due to the atomic memory. Doppler broadening
of the lineshape at 1 mK temperature is included by direct numerical integration of a
one-dimensional Maxwell-Boltzmann velocity distribution.

The two degrees of freedom of the comb, $f_o$ and $f_r$, allow for one mode of the 389 nm comb to
be resonant with the $2\,^3S \rightarrow 3\,^3P_0$ transition and one mode of the 789 nm comb to
be resonant with the $3\,^3P_0 \rightarrow 40\,^3S_1$ transition. In general, there are many
possible solutions for $f_r$ and $f_o$ that satisfy these resonance conditions. For the results
presented in this Section, $f_r$ is chosen to be approximately 94.6 MHz and $f_o$ is -7.6 MHz,
ensuring that the intermediate state is resonant with a mode. It is worth emphasizing that the
general properties of the lineshapes presented in this Section are valid for a comb with a
significantly different $f_r$. For example, a comb at 100 MHz and a peak field of 10$^7$ V/m gives
results very similar to a 500 MHz comb with a peak field of 5$\times$10$^6$ V/m. Because the 389
nm light is generated by frequency doubling the 789 nm light, the two combs share a common $f_r$.
However, the offset frequency $f_{o,2\gamma}$ at 389 nm will be twice that at 789 nm. Therefore, a
change of $\Delta f_o$ at 789 nm causes the detuning from the $40\,^3S_1$ state to change by
$3\Delta f_o$. The peak field strength and transform-limited duration of the 789 nm pulse are
$10^7$ V/m and 30 fs, respectively. To account for the frequency doubling process, the 389 nm
light field strength is reduced to $5 \times 10^6$ V/m and the transform limited pulse length is
increased to 65 fs, reflecting spectral narrowing due to phase matching limitations.  Although for
these simulations the spectral phase is assumed flat, the effect of pulse chirp is small in the
case of multipulse excitation of a two-photon transition with a resonant intermediate state
\cite{Ye06b}.

Unlike traditional Doppler-free spectroscopy with a single cw laser, the wavelengths of the pulses used for the transition under study are quite dissimilar, and they are scanned simultaneously but at differing rates as $f_0$ or $f_r$ is varied. The mismatch in Doppler shifts between the 389 nm and 789 nm light only allows for partial cancellation of the net Doppler shift via absorption from counterpropagating pulses. Figure 2 illustrates the effect of this mismatch on the lineshapes of the 40$^3S_1$ state for various velocity classes, with populations determined by a Maxwell-Boltzmann distribution at 1 mK. For each nonzero velocity class there are four separate peaks, labeled A through D, corresponding to the four possible combinations of resonant photon absorption from counterpropagating and copropagating pulse pairs. The peaks are split because the total residual Doppler shift differs for each of the four combinations.  Referring again to Fig. 2, the peaks detuned furthest for any velocity class, labeled A and D, correspond to the case where both the 389 nm and 789 nm photons are absorbed from a single pulse direction. These peaks are the largest in amplitude and are slightly power broadened, because they arise when both the $2\,^3S_1 \rightarrow 3\,^3P_0$ and $3\,^3P_0 \rightarrow 40\,^3S_1$ transitions are resonant.  This Doppler-shifted resonance condition occurs because $f_o$ for the 389 nm pulse is twice that of the 789 nm pulse, which compensates for the fact that the Doppler shift at 389 nm is almost twice that at 789 nm for any particular atomic velocity. In Fig. 2 the inner two peaks, B and C, are smaller and narrower because they are not exactly resonant with the intermediate state and therefore exhibit less power broadening.

The incoherent sum of many of the lineshapes shown in Fig. 2 with velocities that sample a
Maxwell-Boltzmann distribution results in the final expected lineshape shown in Fig. 3. There are
two clear peaks in this spectrum that arise from the absorption of photons from pulses in both
directions. These two peaks are solely due to the imbalance of Doppler shifts between the 389 nm
and 789 nm light and not due to Autler-Townes splitting, which becomes significant only at
slightly higher field strengths. Figure 4 illustrates the effect of Autler-Townes splitting for a
zero-velocity atom; the gaussian lineshape at the center corresponds to the case discussed thus
far and does not exhibit any splitting. The two other lines shown in Fig. 4 exhibit significant
Autler-Townes splitting of the gaussian lineshape. These two cases correspond to either a ten
times higher electric field for an $f_r$ of 100 MHz, or with similar results, if the field remains
10$^7$ V/m but the repetition frequency is increased tenfold.

It should be noted that the interaction of the counterpropagating pulses with the atoms have a
spatial dependence like the standing wave formed by counterpropagating two single-frequency
lasers, one at 789 nm and the other at 389 nm.  The results presented here are strictly valid only
for the point in space where the counterpropagating pulses perfectly overlap. For positions away
from this, the resonant intermediate state allows for a 389 nm photon to be absorbed from the
left, for example, followed by a 789 nm photon from the right after some time delay. However,
unlike the case of two-photon absorption without a resonant intermediate level, the double-peaked
feature due to the absorption of photons from both of the counterpropagating pulses can be
observed even where the pulses do not overlap spatially owing to the finite lifetime of the
intermediate state.

\begin{figure}
\centering \vskip 0 mm
\includegraphics[width=0.95\linewidth]{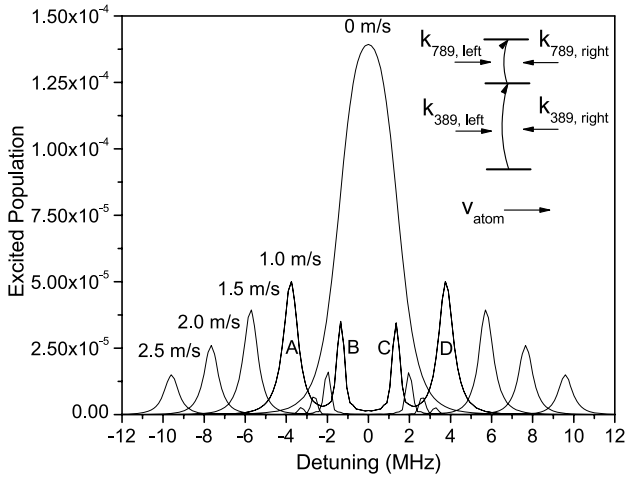}
\caption{\protect\label{fig2} Calculated lineshapes for several velocity groups at constant laser
irradiance, each with four distinct peaks (A-D) from the four possible combinations of counterpropagating and copropagating beams.}
\end{figure}

\begin{figure}
\centering \vskip 0 mm
\includegraphics[width=0.95\linewidth]{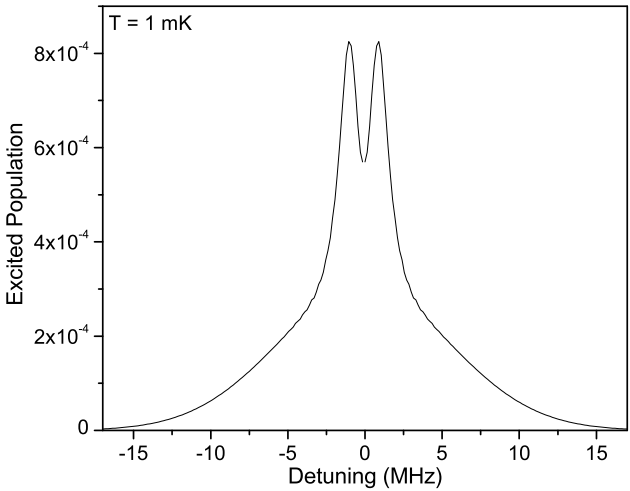}
\caption{\protect\label{fig3} Lineshape with counterpropagating beams for $f_r\approx 100$ MHz and
a peak electric field of $10^7$~V/m, showing full Doppler pedestal, reduced-Doppler peak, and
central sub-Doppler dip.}
\end{figure}

\begin{figure}
\centering \vskip 0 mm
\includegraphics[width=0.95\linewidth]{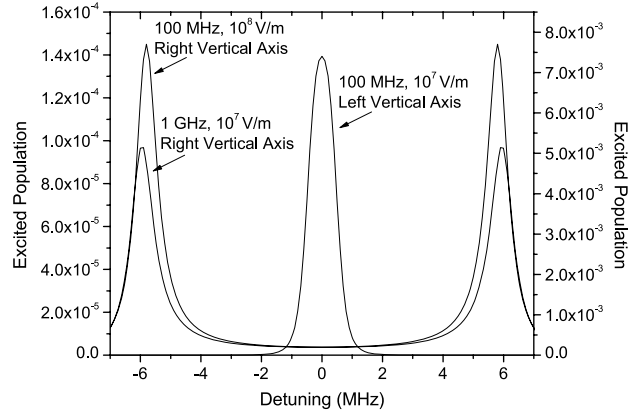}
\caption{\protect\label{fig4} Autler-Townes type splittings for an atom at zero velocity.  Note
that the field and $f_r$ used in Figs. 2 and 3 is too small to exhibit these splittings.}
\end{figure}

To achieve a practical measurement of a series of triplet $ns$ and $nd$ Rydberg states, and thus
to determine the IE of the $2\,^3S_1$ state, an important consideration is to ensure that each
Rydberg state can be excited in isolation and its spectrum unambiguously identified. This requires
a careful selection of the comb parameters $f_o$ and $f_r$ to avoid overlapping excitation of
multiple states by different comb modes. The fundamental comb frequencies centered at 789 nm can
be written as $\nu_n= f_0 + nf_r$, and the second-harmonic comb frequencies near 389 nm as  $\nu_m
= 2f_o + mf_r$.  Using these comb frequencies and the frequencies of the $2\,^3S_1-3\,^3P_0$ and
$3\,^3P_0-(n\,^3S_1, n^3D_1)$ transitions provided by Martin \cite{Martin87}, we employ a search
algorithm to determine whether there are combinations of the parameters \(f_{r}\) and \(f_{o}\)
that excite only a single Rydberg state. This algorithm searches for combinations of \(f_{r}\) and
\(f_{o}\) that place a second-harmonic comb mode $\nu_m$ on resonance with the $2\,^3S_1
\rightarrow 3\,^3P_0$ transition while only one fundamental comb mode $\nu_n$ is on resonance with
a single Rydberg state. We require that all other fundamental comb modes are detuned from the
remaining Rydberg states by at least three Doppler linewidths. Based on the predicted 4 MHz
Doppler-limited linewidth for counterpropagating-beam excitation at 1 mK, it can be determined
that a frequency comb with \(f_{r}>465\) MHz is required to measure all Rydberg states with
principal quantum numbers $7 < n < 40$.  For principal quantum numbers $n > 40$, the $^3S_1$ and
$^3D_1$ Rydberg series are sufficiently congested that extra spectral filtering of the comb is
required.

\section{Intercombination transitions and production of ultracold singlet-state helium}
\label{sec:Intercombination}

The present status of the singlet-state spectrum is somewhat different than for the triplets. The ground $1\,^1S$ state is a special case because it lies about 20 eV below the excited states, and will be discussed separately in Section \ref{sec:Singlets}.  From the metastable $2\,^1S$ state, transitions to high $nD$ Rydberg states with $n$=7-20 have been accurately measured by the groups of Lichten and Sansonetti \cite{Lichten91,Sansonetti92}, yielding an extrapolated IE accurate to $1.5~\times~10^{-10}$, or 0.15 MHz. This result is in reasonably good agreement with theory \cite{Pachucki06c}, and considerably more accurate, so it is not an immediate priority for new experimental work. On the other hand, an accurate measurement of a singlet-triplet intercombination transition between low-$n$ states could extend this same accuracy to the triplet states, providing an independent check on the IE determination proposed in Section \ref{sec:triplet}. Surprisingly, few other singlet states with $n$=2-4 have been accurately measured, and here the theory is well ahead of experiments.  For the $2^1P$ state there is a significant disagreement between experiment and new theoretical calculations by Pachucki \cite{Pachucki06c}.  Assuming that an optical trap can be loaded with singlet metastables, transitions to excited singlet states could be measured in much the same fashion as the triplet spectra discussed in Section \ref{sec:triplet}.  We do not specifically address these measurements here because of this close similarity to the triplet-state experiments.

Instead, the primary emphasis in this Section will be on singlet-triplet intercombination
transitions, which are very difficult to access by conventional room-temperature spectroscopy.  We
start by describing two possible schemes for exploiting intercombination lines to prepare samples
of ultracold singlet-state helium in the $1\,^1S$ state suitable for further study.  We then
discuss a Raman transfer scheme to produce atoms in the singlet metastable state, $2\,^1S$.
Finally, we propose a DFCS measurement of the $2\,^3S-2\,^1S$ interval that takes advantage of
some of these same transitions.

\subsection{Production of ground-state $1\,^1S$ atoms by intercombination transitions}
\label{sec:Intercombination_ground-state}

Unfortunately ground-state helium atoms cannot be cooled directly by any lasers likely to be
available in the near future, and the metastable $2\,^1S$ state has no cycling transitions
suitable for laser cooling.  However, the singlet spectrum can still be accessed indirectly by
starting with a helium MOT in the $2\,^3S$ state and driving transitions to states with mixed
singlet-triplet character.

One prospect for preparing the ground $1\,^1S$ state is the single-photon $2\,^3S \rightarrow 2\,^1P_1$ intercombination transition and the subsequent radiative decay $2\,^1P \rightarrow 1\,^1S$.  Observation of this transition would also allow a new experimental determination of the $2\,^1P$ term energy.  The $2\,^3S - 2\,^1P_1$ oscillator strength has been calculated \cite{Lin77}, and the corresponding dipole matrix element is a few thousand times smaller than for a typical allowed transition:
\begin{equation} \label{D13}
D_{13} = \left\langle {2\,^1P_1} \right|e\textrm{\textbf{r}}\left| {2\,^3S_1} \right\rangle  =
5.65 \times 10^{-33} \ {\rm{C}} \cdot {\rm{m}}.
\end{equation}

This is still large enough to easily allow excitation in a MOT.  If excited resonantly
by a cw laser at 887 nm with irradiance $I$, the excitation rate in the low-power limit is
\begin{equation}
R = \frac{{\Omega^2 }}{\gamma_{2p}}{\rm{,\ where}}~\Omega = \sqrt {\frac{{2I\,D_{13} ^2
}}{{\varepsilon _0 c\hbar ^2 }}} \,.
\end{equation}
An irradiance of $10^4$ W/cm$^2$, easily attained by focusing a 1 W Ti:Sapphire laser to a diameter of about 0.1 mm, will provide a rate of $1.2 \times 10^5$ s$^{-1}$ per atom.  The $2\,^1P$ state then radiates to the ground state by emitting a 58 nm photon at a rate $\gamma_{2p} = 1.8\times10^9$~s$^{-1}$ \cite{NIST}.  Thus the illuminated region is effectively transferred to the ground state in about 10 $\mu$s.  Unfortunately there is appreciable heating due to photon recoil.  The atomic velocity increment from emitting a 58 nm photon is 1.7 m/s, nearly as large as the most probable velocity of 2 m/s for atoms in a 1 mK MOT.  In Section \ref{sec:Singlets} we address the issues associated with loading an optical trap with ground-state atoms under these conditions.

Another possibility for singlet-state production is two-photon excitation of $2\,^3S$ atoms to an
$n\,^1D_2$ state, which is weakly allowed due to spin-orbit coupling with the corresponding
$n\,^3D_2$ state.  For high-$n$ states the coupling due to magnetic fine-structure interactions
\cite{Palfrey83} and the known energy exchange splitting \cite{Farley} can be used to estimate the
$^{1,3}D_2$ mixing.  The resulting wave function admixture, approximately 1.8\% at $n=10$, should
be nearly independent of $n$ because both the exchange splitting and the spin-orbit interaction
scale approximately as $1/n^3$. This is confirmed by an experimental determination that the
$3\,^1D-3\,^3D$ admixture is 1.53\% \cite{Fujimoto86}.  The $3\,^1D$ state is particularly
convenient because it can readily be excited by a combination of the MOT trapping laser at 1083 nm
and an added laser at 588 nm, via the near-resonant pathway $2\,^3S \rightarrow 2\,^3P \rightarrow
3\,^1D_2$.  The atoms will subsequently decay radiatively into the ground $1\,^1S$ state,
primarily by the path $3\,^1D_2 \rightarrow 2\,^1P \rightarrow 1\,^1S$.  The effects of photon
recoil are nearly the same as for the other scheme, since the first decay step emits a near-IR
photon that has little impact.

The two-photon rate can be estimated using a generalization of Eq. \ref{2-photon_Rabi},
\begin{equation} \label{2-color_Rabi}
\Omega_{2\gamma} = \frac{D_{3\,^1\!D-2\,^3\!P} D_{2\,^3\!P-2\,^3\!S}}
{\varepsilon_0 c\hbar ^2 }\frac{\sqrt {I_1 I_2}}{\Delta }.
\end{equation}
If the MOT laser has an irradiance $I_1 = 0.02$ W/cm$^2$ and is detuned from resonance by 20
natural linewidths, and the 588 nm transfer laser has an irradiance $I_2$ = 1 W/cm$^2$, the
two-photon Rabi frequency is $\Omega_{2\gamma} = 1.4 \times 10^7$~s$^{-1}$.  The corresponding
transition rate to the $3\,^1D$ state is very large, $R = 3 \times 10^6$~s$^{-1}$.  This transfer
scheme is much easier to saturate than the single-photon excitation to $2\,^1P_1$
because the stronger singlet-triplet mixing of the $nD_2$ states more than compensates for the
slightly off-resonant intermediate $2\,^3P$ state.

\subsection{Production and trapping of helium in the metastable singlet $2\,^1S$ state}
\label{sec:Intercombination_trapping}

The metastable $2\,^1S$ state has a lifetime of 1/51 s \cite{Lin77} and is a good candidate for
optical trapping if it can be produced at ultracold temperatures.  This would open the way for
precision measurements and many other experiments on the singlet system.  One obvious possibility
is to start with triplet-state helium in a MOT or optical trap and perform a lambda-type
stimulated Raman transfer via the $2\,^1P_1$ state, $2\,^3S \rightarrow 2\,^1P_1 \rightarrow
2\,^1S$.  Assuming the transfer is near-resonant with the $2\,^1P$ state, the required wavelengths
are near 887 nm and 2059 nm.  Unlike the ground-state case, the photon recoil velocity is only
0.064~m/s and contributes no significant heating.

Unfortunately, the $2\,^1S$ state is very susceptible to losses due to inelastic Raman scattering that transfers atoms to the $1\,^1S$ ground state, in this case via absorption at 2059 nm and spontaneous emission at 58 nm.  The cross-section for spontaneous Raman scattering with absorption at $\omega$ and emission at $\omega_{scatt}$ can be estimated using the angular average of the appropriate
terms in the Kramers-Heisenberg formula \cite{Loudon00},
\begin{equation} \label{RamanScatt}
\sigma(\omega) = \frac{{\omega \,(\omega_{scatt})^3 }}{{18\pi \varepsilon _0 \hbar ^2 c^4 }}
\left[ {\sum\limits_n {\frac{{D_{1snp} D_{np2s} }}
{{\omega _{np}  - \omega _{2s}  - \omega }}} } \right]^2.
\end{equation}
We have evaluated this cross-section by summing over intermediate $n\,^1P$ states up to $n=10$. We
determine the magnitudes of the matrix elements from the oscillator strengths of Ref. \cite{NIST}
and their signs from an approximate numerical calculation using phase-shifted Coulomb wave
functions.

For a laser irradiance $I$, the inelastic scattering rate is $R=\sigma I/\hbar \omega$
per atom.  These Raman scattering rates are anomalously large near the $2\,^1S - 2\,^1P$
transition because of the combination of a large oscillator strength and a very high frequency
$\omega_{scatt}$ for decay at 58 nm.  As an example, if the 2059 nm laser is detuned from
resonance by 1 nm and has an irradiance of just 10 W/cm$^2$, the inelastic Raman rate is $2.8
\times 10^4$~s$^{-1}$.  Because the two-photon Rabi frequency must be large enough to compete with
the Doppler width due to the $2\,^1S - 2\,^3S$ energy difference (0.69 MHz at 100 $\mu$K), it is
not possible to obtain efficient Raman transfer in a normal thermal sample without large
scattering losses.  The situation does not improve with increased detuning $\Delta$ from the
$2\,^1P$ state, because the power requirements for efficient two-photon transfer increase as
$\Delta^2$, canceling the decrease of the Raman rates with $1/\Delta^2$.  Although the situation
might be improved considerably using stimulated Raman adiabatic passage (STIRAP), great care would
be required to make the scheme feasible.  The outlook would be much more favorable in a BEC or in
the Doppler-free ``magic wavelength'' optical lattice trap described in Sections \ref{sec:Intercombination_DFCS} and \ref{sec:Singlets_1s2s}.

A more promising approach for ordinary thermal samples is to use a four-photon transfer via the
$3\,^1D$ state, as shown in the center portion of Fig. \ref{He_Levels}.  Only two dedicated lasers
are needed, at 588 nm and 1009 nm, since the MOT trapping radiation at 1083 nm can be used for the
first step, and two equal 1009 nm photons can be used to drive the stimulated $3\,^1D \rightarrow
2\,^1S$ transition. It is best to avoid exact resonance with any of the intermediate states so
that they are not populated appreciably.  The approximate four-photon Rabi frequency is given by a
generalization of Eq. \ref{2-photon_Rabi},
\begin{eqnarray} \label{4-photon_Rabi}
\Omega_{4\gamma} = &&\frac{{D_{2^1\!S - 2^1\!P} D_{2^1\!P - 3^1\!D}
D_{3 ^1\!D - 2^3\!P} D_{2^3\!P - 2^3\!S} }}
{{2\varepsilon _0 ^2 c^2 \hbar ^4 }} \nonumber\\*
&&\times \frac{{\sqrt {I_1 I_2 I_3 I_4 } }}{{\Delta _1 \Delta _2 \Delta _3 }}.
\end{eqnarray}
Here $I_1$, $I_2$ and $I_3$ are the irradiances at 1083 nm, 588 nm, and 1009 nm, respectively, and
the detunings $\Delta_1$, $\Delta_2$, and $\Delta_3$ refer to the $2\,^3P$, $2\,^1D$ and $2\,^1P$
states.  The required matrix elements $D_{ij}$ are obtained from the oscillator strengths in Ref.
\cite{NIST}, together with the singlet-triplet mixing fraction described in Section \ref{sec:Intercombination_ground-state}.

The irradiances and the adjustable detunings $\Delta_1$ and $\Delta_2$ are optimized to give a
large transfer rate without significant inelastic Raman scattering either from the $2\,^1S$ state
or from atoms that are excited to the $2\,^3P$ state by the slightly off-resonant 1083 nm MOT
laser.  With $\Delta_1 = 2\pi \times$ 25 MHz, $\Delta_2 = 2\pi \times$ 1 GHz, $I_1 =$~0.01
W/cm$^2$, $I_2 = 5 \times 10^3$ W/cm$^2$, and $I_3 = 3 \times 10^5$ W/cm$^2$, the Rabi frequency
is $\Omega_{4\gamma} = 3.1 \times 10^5$~s$^{-1}$.  The high irradiance at 1009 nm should be easily
attainable with a tightly focused laser, given the availability of 1 W tapered amplifiers at this
wavelength \cite{Toptica}.  With this arrangement all of the atoms will be transferred in about 1
microsecond, so the MOT can be turned off without loss due to thermal motion.  Further, the power
broadening and transition-time broadening match the transition width well with the Doppler width
at 100 $\mu$K.  We estimate that Raman scattering losses will be only about 1\%.

The ultracold $2\,^1S$ atoms must be trapped if useful experiments are to be performed on them.
An optical trap can be constructed in either of two ways:

(1) The $2\,^1S$ polarizability is small but positive at 588 nm \cite{Chen95}.  If the laser is
sufficiently strong it can serve a dual role: after driving the stimulated decay to the metastable
state, it can also act as a far-off-resonance trap (FORT) if the irradiance is increased to
$\gtrsim 10^6$ W/cm$^2$.  The detuning $\Delta_2$ for the transfer scheme can be increased so that
the transfer and scattering rates are unchanged.  The trapping time will be limited, however, by
inelastic scattering of FORT light by the $2\,^1S$ atoms: for a trap depth of 100 $\mu$K, the
scattering rate is about 10$^3$ s$^{-1}$ per atom.

(2) If longer trapping times are desired, a 10.6 $\mu$m CO$_2$ laser can be substituted.  A trap depth of 100 $\mu$K is attained at $5 \times 10^8$ W/cm$^2$ and the inelastic scattering rate is 44 s$^{-1}$, slightly less than the radiative decay rate of 51 s$^{-1}$.

\subsection{DFCS measurement of the $2\,^1S-2\,^3S$ interval}
\label{sec:Intercombination_DFCS}

It was recently proposed by van Leeuwen and Vassen that the doubly-forbidden $2\,^3S-2\,^1S$
transition could be measured by using a 1.557 $\mu$m laser to directly drive the very weak M1
transition between these two levels \cite{Vassen06}. The first of the Raman-type schemes proposed
in the preceding Section suggests an alternative method: two-photon DFCS of the $2\,^3S
\rightarrow 2\,^1P_1 \rightarrow 2\,^1S$ transition can provide a self-calibrating measurement.
Even though this scheme is problematic for efficient transfer to the singlet state, it is much
better suited to spectroscopy, where high transfer efficiencies are not required, and Raman
scattering losses might even be exploited as a detection mechanism.  It is probably not feasible
to use a Ti:Sapphire based comb directly, since the comb cuts off at about 1.2 $\mu$m and this is
too far from the singlet resonance at 2.06 $\mu$m to provide suitable two-photon rates.  However,
a fiber-laser based frequency comb could be used together with its second harmonic to provide
frequency-domain teeth near both 887 nm and 2.06 $\mu$m.

It would be necessary to avoid near-resonance with the intermediate $2\,^1P$ state, thereby controlling losses from inelastic Raman scattering while also avoiding the complex lineshapes encountered in the simulations of triplet spectra described in Section \ref{sec:triplet_Rydberg}.  If spectrally filtered combs are used with an average detuning of, say, 10$^{12}$ Hz, the transition can be treated as a quasi-cw process just as for the two-photon excitation discussed in Section \ref{sec:triplet_two-photon}. Based on Eq. (\ref{2-color_Rabi}), a power of 50 mW focused to a radius of $\omega_0 = 50\ \mu$m would yield a Rabi frequency $\Omega_{2\gamma} = 10^4$~s$^{-1}$, assuming the use of copropagating beams with the trap turned off.  This is easily enough to yield a detectable signal. The signal can be monitored either by depletion of the triplet metastables or by detection of the singlet
metastables.

This method differs greatly in both physics and methodology from the proposal of Vassen's group,
and would provide a valuable independent determination of the singlet-triplet intercombination
energy.  The principal limitations on the linewidth, though, are similar to those mentioned by
them: Doppler broadening if a normal thermal sample is used, or light shifts if a tight optical
trap is used to reduce the Doppler shift by Lamb-Dicke narrowing.  Since the thermal Doppler width
can be reduced to 700 kHz or less, and the frequency measurements are self-calibrating, an
absolute accuracy of 10 kHz or better should be attainable in a normal thermal sample.

Even higher accuracies could be attained by use of an optical lattice trap in which the $2\,^1S$ and $2\,^3S$ states experience identical optical potentials, a so-called ``magic wavelength'' trap \cite{Ido03,Boyd06}.  The trap can be left on continuously during measurements because the trap level spacings are identical, and the trap shifts cancel.  It is mentioned in Ref. \cite{Vassen06} that the $2\,^1S$ and $2\,^3S$ polarizabilities are equal near 410 nm, and our own calculation in Section \ref{sec:Singlets_1s2s} indicates a similar magic wavelength of 412 nm.  Even though the polarizabilities at this wavelength are small, ~5-10 $a_0^3$, and the wavelength is not convenient for high-power cw lasers, usable trap depths can be attained.  A depth of about 70-150 $\mu$K can be achieved using a resonant optical cavity with an average circulating power of 5 W and a gaussian beam waist $\omega_0 = 10 \mu$m.  Under these conditions the predicted inelastic Raman scattering rate of trap light by a $2\,^1S$ atom is 916 s$^{-1}$.  The $2\,^3S - 2\,^1S$ transition should be in the Lamb-Dicke regime in which the Doppler and recoil shifts become negligibly small.  This topic is discussed further in Section \ref{sec:Singlets_1s2s}.  With no inherent limits apart from the shortened singlet-state lifetime due to Raman scattering, the linewidth would be about 150 Hz.

\section{Far-UV DFCS Spectroscopy of the singlet states}
\label{sec:Singlets}

\subsection{VUV frequency comb excitation of $1\,^1S \rightarrow n\,^1P$ transitions}
\label{sec:Singlets_VUV}

If a far-UV frequency comb is available near 51 nm, it is straightforward to produce singlet
ground-state atoms by the ``dumping" scheme described above, and then to directly excite Rydberg
$n\,^1P$ states via single-photon excitation by an individual frequency-domain comb tooth.  The
excitation rate will be the same as for a cw laser with the same time-averaged irradiance as the
resonant tooth of the frequency comb.  The principal limitation is the average Doppler shift of
about 23 MHz due to the photon recoil velocity acquired during decay to the $1\,^1S$ state, $v_r
\simeq 2$~m/s.  This velocity also limits the interaction time in the far-UV laser beam.  However,
because this Doppler profile is still narrow compared to the frequency comb spacing $f_r$ of ~500
MHz, we do not expect complications such as the complex lineshapes arising from velocity-selective
optical pumping that have been observed in room-temperature gases \cite{Pichler05,Pichler07}.

We estimate the electric dipole matrix elements $D(n) = \left\langle n\,^1P\right|er\left|1\,^1
S\right\rangle$ by using the known $A$ coefficient for $n$=10 (from Ref. \cite{NIST}) to find
$D(10)$, then scaling by $(n^*)^{-3}$, where $n^*$ is the effective principal quantum number.  We also
assume approximate $(n^*)^{-3}$ scaling for the lifetimes $\tau(n)$, using as a reference point the
$21P$ lifetime calculated by Theodosiou \cite{Theodosiou84}.  This yields a natural linewidth of
12 MHz for $n=6$ and just 0.042 MHz for $n=40$, both much smaller than the Doppler width of about
20 MHz at 100 $\mu$K or the similarly sized Doppler shift due to photon recoil in production of
the ground-state atom.  Thus inhomogeneous Doppler broadening dominates the linewidth.

The Rabi frequency for excitation by a single tooth of the comb with irradiance $I_t$ is given by
\begin{equation}
\Omega(n) = \sqrt{\frac{{2I_t D(n)^2 }}{{\varepsilon _0 c\hbar ^2 }}},
\end{equation}
and the steady-state probability of excitation at a detuning of $\delta$ is given by the usual cw
expression,
\begin{equation}
P(n) = \frac{\Omega^2 (n)}{4}\frac{1}{{\delta^2 + \left({\frac{1}{{2\tau(n)}}} \right)^2 +
\frac{{\Omega^2 (n)}}{2}}}.
\end{equation}
We assume a far-UV frequency comb with a total time-averaged power of $10^{-8}$ W in the harmonic
near 51 nm, with $f_r=500$~MHz and a bandwidth of $10^{12}$~Hz.  If focused to a 1/$e^2$ radius of
35 $\mu$m, the on-resonance excitation rate per atom is 2.6~s$^{-1}$, independent of $n$.  Taking
into account the inhomogeneous broadening and assuming the ground-state atom distribution matches
the MOT parameters listed in Section \ref{sec:DFCS}, the total excitation rate is roughly 140 atoms/s at
$n$=40, or $4 \times 10^4$ atoms/s at $n$=6.

These rates are adequate if delayed Stark field ionization is used for high-$n$ states, or an
auxiliary ionizing laser at lower $n$.  The accuracy will be limited by the Doppler width and the
limited signal sizes; a reasonable guess is that 1 MHz accuracy should be attainable.

\subsection{Optical trapping of $1\,^1S$ helium atoms} \label{sec:Singlets_trapping}

For precision spectroscopy of the $1\,^1S - 2\,^1S$ interval it is necessary to optically trap
atoms in the $1\,^1S$ ground state after producing them with one of the ``dumping'' schemes in
Section \ref{sec:Intercombination_trapping}. This is unusually difficult because of its extremely small dc polarizability
$\alpha=1.38\ a_0^3$ \cite{Thomas72}, which is nearly constant throughout the optical and near-UV
regions \cite{Starace03}.  Generally a resonant buildup cavity would be needed to provide
sufficient irradiance.  For example, a cavity with a gaussian waist radius of $\omega_0 = 20 \mu$m
and an average circulating power of 80 W yields a trap depth of about 80 $\mu K$.

The loading efficiency is limited primarily by the broad speed distribution of the ground-state atoms, which experience randomly directed velocity increments of 1.71 m/s due to photon recoil from the 58 nm decay radiation that produces them.  We have devised a simple Monte Carlo model to determine the fraction of ground-state helium atoms that can be trapped, by assuming an initial temperature $T_i$ and then assigning to each atom a random recoil direction. We then determine the trappable fraction by comparing the final kinetic energy to the depth of the ground-state optical trap.  Table \ref{tab:TrappedFraction} lists the trappable fractions for various trap depths for two values of $T_i$.  As $T_i$ is increased, the number of very slow atoms increases significantly, rather than declining as it would in a normal thermal sample.  This occurs because at lower temperatures the recoil velocity becomes large compared to the initial thermal velocity spread, leaving very few atoms near zero velocity.

\begin{table}
\caption{Fraction of $1\,^1S$ atoms that can be optically trapped, taking into account a 58 nm
photon recoil, for two values of the temperature $T_i$ prior to the ``dump'' step.}
\label{tab:TrappedFraction}
\begin{tabular}{lll}
\hline\noalign{\smallskip}
Trap depth ($\mu$K)&$T_i = 1$ mK&$T_i = 100 \mu$K\\
\noalign{\smallskip}\hline\noalign{\smallskip}
100 & 0.012 & 0.0029\\
200 & 0.031 & 0.017\\
250 & 0.044 & 0.032\\
300 & 0.056 & 0.052\\
500 & 0.114 & 0.194\\
1000 & 0.27 & 0.69\\
\noalign{\smallskip}\hline
\end{tabular}
\end{table}

Assuming that the triplet metastable atoms are initially loaded into a dense optical trap, it
should be feasible to work with trappable fractions as low as 0.01.  This can still yield hundreds
or thousands of optically-trapped ground-state atoms, sufficient to allow DFCS spectroscopy.

\subsection{Measuring the $1\,^1S \rightarrow 2\,^1S$ transition in a ``magic wavelength" optical
lattice trap}
\label{sec:Singlets_1s2s}

\begin{figure}
\centering \vskip 0 mm
\includegraphics[width=0.95\linewidth]{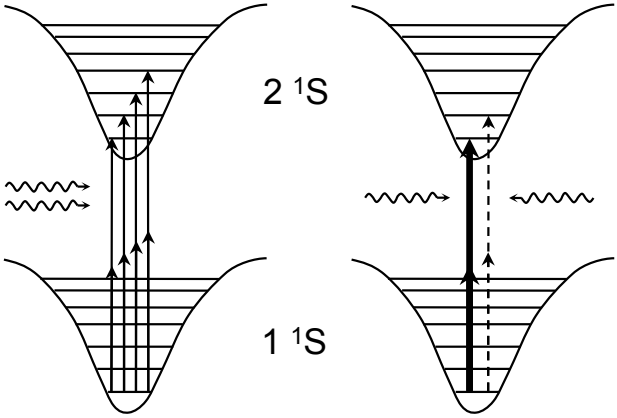}
\caption{\protect\label{MagicTrap} Schematic representation of two-photon transitions
in a magic wavelength trap, for copropagating (left) and counterpropagating (right)
configurations.}
\end{figure}

The narrow 8 Hz natural linewidth of the 2 $^1S$ state makes it a natural candidate for two-photon excitation directly from the ground state.  Doppler-free excitation is possible using counterpropagating 120 nm VUV lasers, and has already been used for a measurement accurate to 45 MHz using nanosecond lasers \cite{Bergeson98,Bergeson00,Eikema97}.  Unfortunately, the low power of currently available VUV frequency combs, combined with the 1.7 m/s recoil velocity of ground-state atoms produced by the schemes proposed in Section \ref{sec:Intercombination_ground-state}, makes a DFCS experiment challenging.  One possibility is to amplify a small number of time-domain pulses from the comb to perform Ramsey-type spectroscopy, an approach being investigated by the group of Eikema \cite{Eikema06}.

\begin{figure}
\centering \vskip 0 mm
\includegraphics[width=0.95\linewidth]{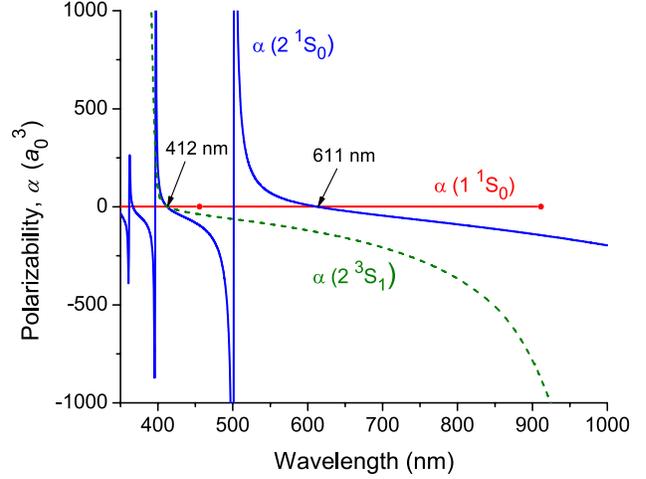}
\caption{\protect\label{MagicLambda} (color online) Polarizabilities of the $1\,^1S$ state (red) from Ref. \cite{Starace03}, the $2\,^1S$ state (blue) from our own calculations, and the $2\,^3S$ state (dashed green) from Ref. \cite{Chen95}.  The red $1\,^1S$ curve is nearly constant at 1.39 $a_0{}^3$.  Magic wavelengths occur where the $1\,^1S$ and $2\,^1S$ curves intersect, at 611 nm, 412 nm, 366 nm, etc. At the ``triple magic'' wavelength of 412 nm the polarizabilities of all three states are similar, small, and positive.}
\end{figure}

Here we examine an approach that would retain the full resolution of DFCS with a continuous pulse
train, two-photon DFCS of the $1\,^1S \rightarrow 2\,^1S$ transition in a magic wavelength optical
lattice.  The $1\,^1S$ and $2\,^1S$ states experience identical optical potentials as shown in
Fig. \ref{MagicTrap}.  The $2\,^1S$ polarizability has been calculated by Chen \cite{Chen95} from
dc to 506 nm, and in this region there is one magic wavelength near 611 nm.  Here the
polarizability crosses through zero because the photon energy is between the $2\,^1P$ and $3\,^1P$
states.  To search for additional magic wavelengths we have performed an approximate calculation
of the polarizability over an extended range, by considering the effects of singlet $nP$ states up
to $n=10$:
\begin{equation}
\alpha_{2^1\!S} (\omega ) \simeq \frac{2}{{3\hbar }}\sum\limits_{n = 2}^{10}
{\frac{{\left( {D_{n^1\!P - 2^1\!S} } \right)^2
\left( {\omega _{n^1\!P}  - \omega _{2^1\!S} } \right)}}
{{\left( {\omega _{n^1\!P}  - \omega _{2^1\! S} } \right)^2  - \omega ^2 }}} + k.
\end{equation}
The data of Ref. \cite{NIST} are used for the atomic energy levels and oscillator strengths.  A small additive constant $k = 8.2\ a_0^3$ is used to adjust the 611 nm magic wavelength into exact coincidence with Ref. \cite {Chen95}; with this adjustment the agreement is within about 2 $a_0^3$ over the entire range of Ref. \cite{Chen95}.  The results are shown in Fig. \ref{MagicLambda} together with the polarizabilities of the $1\,^1S$ and $2\,^3S$ states.  Magic wavelengths are apparent at 611 nm, 412 nm, 366 nm, and various shorter wavelengths.

Remarkably, a ``triple magic wavelength'' is predicted near 412 nm, where all three states ($1\,^1S$, $2\,^1S$, and $2\,^3S$) have nearly the same polarizability, which is small and positive. Any pair of these states can be exactly matched by making very small adjustments to the wavelength.   Although the 611 nm wavelength is more convenient for cw lasers, the ability to trap all three states simultaneously at 412 nm may make it a better choice.

Table \ref{tab:MagicTrap} summarizes the properties of a 412 nm trap produced by a resonant cavity with a 20 $\mu$m beam waist and a spatially averaged intracavity power of 80 W.  The peak power at the antinodes is twice as high.  It is assumed that the seventh-harmonic comb at 120 nm has a fractional bandwidth of 5 \%.  The lattice trap is strongly quantized along the cavity axis, comprising only a few quantum levels depending on the choice of laser wavelength and irradiance.  While the trap is stable for $1\,^1S$ atoms, it causes losses for $2\,^1S$ atoms due to inelastic Raman scattering.  The loss rate in Table \ref{tab:MagicTrap} is calculated from Eq. (\ref{RamanScatt}) at 1/2 of the peak irradiance, as an approximation to the average potential experienced by the atoms as they move about in the trap.  This loss rate increases linearly with the trapping potential, and if a 611 nm trap is used rather than 412 nm the rate is increased by a factor of 2.6 for a given trap depth.

\begin{table}
\caption{Parameters and trap properties for $1\,^1S \rightarrow 2\,^1S$ transitions with counterpropagating beams in a magic wavelength optical lattice trap at 412 nm.  Residual Doppler width is for an unconfined sample at 80 $\mu$K.  Properties for a single-photon 120 nm transition are shown for comparison only.}
\label{tab:MagicTrap}       
\begin{tabular}{lll}
\hline\noalign{\smallskip}
Parameter&Value&Notes  \\
\noalign{\smallskip}\hline\noalign{\smallskip}
$\lambda$ & 412 nm&\\
$P_{\textrm{avg}}$&80 W&in resonant cavity\\
$\alpha$ & 1.39 a.u.&\\
$\omega_0$ & 20 $\mu$m\\
$U_0$ & 80 $\mu$K&$=10.4 \times 10^6$~s$^{-1}$\\
$\omega_{\textrm{axial}}$ & $6.2 \times 10^6$~s$^{-1}$&In harmonic approx.\\
$\omega_{\textrm{Doppler}}$ & $3.7 \times 10^6$ s$^{-1}$&For 5\% VUV bandwidth\\
$\omega_{\textrm{recoil, resid}} $& $0.12 \times 10^6$~s$^{-1}$&Two-photon, 5\% bandwidth\\
$R_\textrm{scatt}$&$1.83 \times 10^3$~s$^{-1}$&For $2\,^1S$ at 1/2 $U_0$\\
$\eta_{\textrm{Lamb-Dicke}}$& 0.14&Two-photon, 5\% bandwidth\\
$\eta_{\textrm{one-photon}}$& 1.87&For one 120 nm photon\\
$\omega_{\textrm{recoil, 120}} $& $21.8 \times 10^6$~s$^{-1}$&For one 120 nm photon\\
\noalign{\smallskip}\hline
\end{tabular}
\end{table}

A possible experimental scheme for a $1\,^1S - 2 \,^1S$ measurement in this trap is as follows:

1. Form a FORT for the triplet metastable state.  This could be done using a ``triple magic'' trap at 412 nm or with a near-IR laser.   Using a trap depth of about 1 mK, it should be possible to load at least several percent of the MOT atoms (i.e., ~$3 \times 10^6$ atoms) using a molasses cooling cycle.

2. Dump the atoms to the singlet ground state using a laser at either 887 nm or 588 nm, as described in Section \ref{sec:Intercombination_ground-state}.  Trap the ground-state atoms in an optical lattice at 611 or 412 nm.

3. If a relatively deep trap ($\sim$1 mK) is used initially, perform a short evaporation cycle by reducing the depth to about 50-100 $\mu$K.

4. Perform two-photon $1\,^1S \rightarrow 2\,^1S$ excitation with the VUV frequency comb.  The $2\,^1S$ state can be detected with near-unit efficiency by laser ionization as described below.

If this were a single-photon transition at 120 nm there would be little suppression of Doppler and recoil effects, given the Lamb-Dicke parameter of $\eta = 1.87$.  However, the two-photon spectrum comprises two distinct contributions: (1) absorption of a pair of copropagating 120 nm photons is possible throughout the focal volume. Under this scenario, the Lamb-Dicke condition does not hold and the optical lattice trap will not provide discrete spectral features in the two-photon transition spectrum. A broad spectral background is formed.  (2) Absorption of a counter-propagating photon pair is possible only in the region where they spatially overlap, about 30 micron long for pulses 100 fs in duration. However, in this case the absence of recoil establishes a nearly perfect Lamb-Dicke condition. Even if we consider the finite bandwidth of the VUV comb, the recoil shift is still far suppressed, as shown in Table \ref{tab:MagicTrap}, since it depends quadratically on the total photon momentum transferred. \textit{The counter-propagating beam spectrum will comprise a sharp carrier with very weak sidebands at} $\pm \omega_\textrm{axial}$.  We  do not expect to find any Doppler or recoil shifts in the resultant two-photon spectrum.

Under these conditions the linewidth will be limited primarily by the destruction of $2^1S$ atoms by inelastic Raman scattering in the magic-wavelength trap, which has a rate $R_{\textrm{scatt}} = 1830$~s$^{-1}$ that far exceeds the radiative decay rate.  For the parameters shown in Table \ref{tab:MagicTrap} the resulting linewidth is only about 300 Hz.  The two-photon Rabi frequency and the corresponding transition rate are found by replacing the excited-state lifetime with this scattering rate,
\begin{equation}
\Omega _{2\gamma }  = \frac{{e^2 (2I_{\textrm{UV}} )M}}{{\hbar ^2 c\varepsilon _0 }}
\ \ \textrm{and}\ \ R_{2\gamma }  = \frac{{\Omega _{2\gamma } ^2 }}{R_{\textrm{scatt}}},
\end{equation}
where $I_{\textrm{UV}}$ is the the total seventh-harmonic comb irradiance and $M$ is the transition moment defined in Ref. \cite{Bergeson99}.  The factor of two takes into account the counterpropagating-beam configuration.  For $I_{\textrm{UV}} = 2 \times 10^{-6}$~W and a five-micron beam waist, the two-photon rate is $R_{2\gamma} = 8.8$~s$^{-1}$ per atom.  This is quite usable because the trap lifetime for ground-state atoms is limited only by background gas collisions and should exceed 1~s.  Detection can be performed by a cw or rapidly pulsed laser tuned above the $2\,^1S$ ionization limit at 312 nm.  The cross section near threshold is $10^{-17}$ cm$^2$ \cite{Stebbings73}, so a focused laser irradiance of 100 W/cm$^2$ will yield an ionization rate of  1600 s$^{-1}$ that is comparable to the Raman loss rate.

\section{Conclusions}
\label{sec:Conclusions}

The measurements and techniques proposed in this paper would serve several objectives: the development and testing of new methods for self-calibrating DFCS spectroscopy in the UV region, the development of techniques for producing and trapping ultracold singlet-state helium atoms, and most important, the prospect for improving the accuracy of the UV spectrum of helium by several orders of magnitude. Major improvements will require both the use of trapped ultracold atoms and the development of new laser tools that avoid the limitations of nanosecond pulsed lasers.

The first experiment we have described, a measurement of the triplet $2S-4S$ interval, utilizes a relatively simple arrangement that could be accomplished either with cw lasers in an ultracold helium sample or by DFCS, and will serve as a good cross-check on the accuracy of the comb-based measurements. The proposed experiments on higher triplet states will establish an improved IE for the metastable $2\,^3S$ state while also providing a rich testing ground for DFCS, including the prospect of using coherent control techniques to control the excitation of the dense Rydberg-state spectrum.  We also note that ``magic wavelength" optical lattice trapping for the $2\,^3S-2\,^3P$ transition could enable improved measurements using cw lasers.

The proposed methods for producing trapped singlet-state atoms via intercombination transitions should find application for numerous purposes besides those described here. It should also be possible to precisely measure the intercombination spectrum by using a comb to drive two-photon $2\,^3S \rightarrow 2\,^1S$ transitions.  The linewidth could be extremely narrow if a magic wavelength trap at 412 nm were utilized to eliminate Doppler and recoil broadening.

Probably the most exciting possibility is to use a frequency comb in the far-UV region either to excite transitions directly from the ground state to $n\,^1P$ states (fairly straighforward but severely limited by Doppler broadening), or in a magic wavelength trap at 611 nm or the ``triple magic'' wavelength of 412 nm.  The magic wavelength trap would open the way to extremely accurate measurements and to exploring the limits of two-photon DFCS spectroscopy with continuous pulse trains in the far-UV region.  The resulting measurement of the ground-state binding energy would serve as a lasting benchmark for testing
two-electron QED theory.

\section{Acknowledgments}
We thank our colleagues T. Zelevinsky, A. Pe'er, R. J. Jones, K. D. Moll, and T. Ido for useful discussions.  E.E. Eyler thanks JILA for hosting him as a short-term visitor.  We gratefully acknowledge support from AFOSR, DARPA, and NIST for the work performed at JILA, and from NSF for the work at UConn.

%
%

\end{document}